\documentclass[12pt,a4paper]{article}

\usepackage{amsmath,amssymb}
\usepackage{graphicx}
\usepackage[colorlinks=true,linkcolor=blue,citecolor=blue,urlcolor=blue]{hyperref}
\usepackage{bm}
\usepackage{tikz}
\usetikzlibrary{decorations.pathmorphing,calc,arrows.meta}
\usepackage[margin=1in]{geometry}

\begin{document}

\title{Boomeranging through the Earth: when free fall doesn't maximize proper time}

\author{Henrique Gomes\\
\small University of Oxford, Oriel College, Oxford OX1 4EW, United Kingdom\\
\small University of Bonn, Bonn, Germany\\
\small \texttt{gomes.ha@gmail.com}}

\date{\today}

\maketitle
\begin{abstract}
In teaching special relativity, John Bell emphasised the value of pre-relativistic intuitions. His famous spaceship puzzle showed that, in certain cases, the older heuristics of Lorentz and FitzGerald get the right answer to a genuinely relativistic problem more reliably than the principles of special relativity themselves. In teaching general relativity, John Wheeler emphasised the value of simplicity, illustrating curved-spacetime geometry with capsules boomeranging through a tunnel along the Earth's diameter. Here I put the two ideas together. I share a simple thought experiment, involving genuinely general-relativistic effects, in which the standard GR heuristic---``freely falling observers maximise proper time''---gives the wrong answer, while the older heuristic of velocity and gravitational time dilation, applied in a single frame, gets it right immediately. The resolution involves conjugate points and the distinction between local and global extremality of geodesics, but the moral is the same as Bell's: do not let a newer slogan overwrite a perfectly sound older intuition.
\end{abstract}
\section{Introduction}\label{sec:intro}

\begin{quote} ``I have for long thought that if I had the opportunity to teach this subject [special relativity], I would emphasize the continuity with earlier ideas. Usually it is the discontinuity which is stressed, the radical break with more primitive notions of space and time. Often the result is to destroy completely the confidence of the student in perfectly sound and useful concepts already acquired.''\end{quote}
Thus begins Bell's article ``How to Teach Special Relativity''~\cite{Bell1976}. He went on to describe the Dewan--Beran thought experiment~\cite{DewanBeran1959}: two spaceships, connected by a taut thread, undergo identical acceleration programs as measured in an inertial frame~$S$. Does the thread break? Bell reportedly polled his colleagues at CERN and found that opinion was sharply divided. The correct answer---yes, it breaks---is reached most directly by applying the pre-relativistic intuitions of Lorentz and FitzGerald to the material constitution of the thread: the thread Lorentz-contracts, but the inter-ship distance (held fixed by construction in~$S$) does not. The same answer can be extracted from the principles of special relativity alone, but with considerably more fuss. Bell's point was that the older, ``pre-relativistic'' heuristic is, in this case, the more reliable guide to a genuinely relativistic problem.

I want to make the analogous point one level up: there is a simple problem involving general-relativistic effects in which a GR heuristic actively misleads, while an older, pre-geometric heuristic gets the right answer at once.

The GR heuristic in question is the slogan that freely falling observers---those following timelike geodesics---maximise proper time. The older heuristic is the rule that a clock in motion runs slow and a clock lower in a gravitational potential runs slow, both contributions compounding in the obvious way. This rule requires nothing beyond special relativity and the equivalence principle---no Riemannian geometry, no geodesics. I will pit the two against each other.

The setting is borrowed from Wheeler's ``boomeranging through the Earth'' capsule~\cite{Wheeler1990}. Imagine the Earth bored through with a vacuum tube along a diameter. Two observers start at the same point on the surface at the same moment:
\begin{itemize}
    \item \textbf{Alice} jumps into the tube and free-falls through the Earth's centre, out the far side, and back. She follows a timelike geodesic and returns roughly 84~minutes later.
    \item \textbf{Bob} stays put on the surface. He is not in free fall: the Earth pushes him with a proper acceleration~$g$.
\end{itemize}
Who ages more? I invite the reader to commit to an answer before reading on---Bell would have approved.

The GR heuristic answers immediately: Alice is in free fall, Bob is accelerated, so Alice should age more. But the older heuristic points the other way. In the Earth's rest frame, Alice has a high average speed \emph{and} spends much of her trip deeper in the gravitational potential than Bob. Both effects produce time dilation---both reduce her elapsed proper time relative to his. The older heuristic says Bob ages more.

The older heuristic is right. As the weak-field calculation in Sec.~\ref{sec:weakfield} confirms, Bob ages about $1.8\;\mu$s more than Alice per round trip. The GR slogan gives the wrong answer.

It is instructive to bring in a third observer:
\begin{itemize}
    \item \textbf{Charlie} is launched vertically upward at the moment Alice jumps, with just the right velocity to return in free fall at the same reunion event. He, too, follows a timelike geodesic.
\end{itemize}
Now the full ordering is
\begin{equation}
    \tau_{\text{Charlie}} > \tau_{\text{Bob}} > \tau_{\text{Alice}}\,.
    \label{eq:ordering}
\end{equation}
Charlie conforms to the GR slogan (a geodesic observer ages more than an accelerated one), but Alice violates it. The same accelerating observer, Bob, sits on opposite sides of the comparison depending on which geodesic he is paired with. The older heuristic, meanwhile, gets the ordering right in every case: Charlie reaches high altitude and lingers near apogee, gaining on the gravitational term and paying little kinematic penalty; Alice moves fast and sits deep, losing on both counts; Bob, stationary at the surface, falls in between.

So what goes wrong with the GR slogan? The qualification it suppresses is not about exotic matter or non-trivial topology. It is the standard geometric fact that a timelike geodesic maximises proper time only \emph{locally}---only up to its first \emph{conjugate point}. A conjugate point is, by definition, a place where a whole family of geodesics leaving the starting event is refocused; here the relevant family is the set of nearby free-fall worldlines that start at the same event with slightly different initial velocities. Alice's geodesic reaches such a point at the \emph{antipodal surface}---the point on the far side of the Earth, diametrically opposite where she jumped in---where the Earth's mass focuses those neighbouring geodesics back together. (Section~\ref{sec:conjugate} makes this precise.) After that, her geodesic is no longer a proper-time maximiser, and the slogan does not apply.

The geometric notion of conjugate points is well known---it is central to the singularity theorems~\cite{Penrose1965,Hawking1966}. But to my knowledge, no one has used it to exhibit a simple case, involving genuinely general-relativistic effects, in which the GR heuristic actively misleads while the older, pre-geometric heuristic does not.

The plan is as follows. Section~\ref{sec:weakfield} carries out the weak-field calculation. Section~\ref{sec:conjugate} explains conjugate points and develops a Riemannian analogy on the 2-sphere. Section~\ref{sec:discussion} discusses the connection to the standard twin paradox and to the case of non-trivial topology. Section~\ref{sec:conclusion} concludes.

\section{The calculation}\label{sec:weakfield}

Idealise the Earth as non-rotating, spherically symmetric, and of uniform density; assume an evacuated tunnel of negligible width---so that its mass deficit and the attendant departure from spherical symmetry can be ignored---and negligible drag. The whole calculation is weak-field and slow-motion; the quantitative consequences are spelt out after Eq.~\eqref{eq:weakrate} below.

In the standard regime, the metric in the Earth's static rest frame is
\begin{equation}
    ds^2 \simeq -\!\left(1+\frac{2\Phi}{c^2}\right)c^2\,dt^2 + d\ell^2\,,
    \label{eq:weakmetric}
\end{equation}
where $\Phi(\bm{x})$ is the Newtonian gravitational potential, $t$ is the time coordinate of the Earth's static rest frame (coinciding with proper time far from the Earth), $\bm{x}$ is the (Cartesian) spatial position, and $d\ell^2 = \delta_{ij}\,dx^i\,dx^j$ is the flat spatial line element. Equation~\eqref{eq:weakmetric} is the standard weak-field, slow-motion metric of a static Newtonian source, retained to first order in $1/c^2$: outside the Earth it agrees with the weak-field limit of the exterior Schwarzschild metric, and inside with that of Schwarzschild's constant-density interior solution.\footnote{Schwarzschild treated the constant-density sphere in the second of his two 1916 papers~\cite{Schwarzschild1916}.} A fuller expansion would also multiply the spatial part $d\ell^2$ by a factor $1-2\Phi/c^2$; but $d\ell/dt$ is already first order in the small velocity, so this extra factor changes the proper time only at $O(1/c^4)$ and is consistently dropped. For a timelike worldline, the elapsed proper time $\tau$ obeys $d\tau^2 = -ds^2/c^2$. Substituting Eq.~\eqref{eq:weakmetric} and writing $v = |d\bm{x}/dt|$ for the coordinate speed,
\begin{equation*}
    \left(\frac{d\tau}{dt}\right)^{\!2} = 1 + \frac{2\Phi}{c^2} - \frac{v^2}{c^2}.
\end{equation*}
Expanding the square root to first order in $\Phi/c^2$ and $v^2/c^2$ gives
\begin{equation}
    \frac{d\tau}{dt} \simeq 1 + \frac{\Phi}{c^2} - \frac{v^2}{2c^2}\,.
    \label{eq:weakrate}
\end{equation}
This is the only relativistic input we need. The two correction terms---gravitational and kinematic time dilation---are the ``perfectly sound and useful concepts'' that do the work.

The trajectories themselves come, at the same order, from the spatial part of the timelike geodesic equation for the metric~\eqref{eq:weakmetric}, which in the slow-motion limit reduces to Newton's equation\footnote{The reader may confirm this directly: for the metric~\eqref{eq:weakmetric} the only connection coefficient surviving at this order is $\Gamma^i{}_{00}=\partial^i\Phi$, so the spatial part of the geodesic equation $\frac{D}{d\tau}\frac{dx^\mu}{d\tau}=0$ becomes $\ddot{x}^i=-\partial^i\Phi$.}
\begin{equation}
    \frac{d^2 x^i}{dt^2} = -\partial^i \Phi.
    \label{eq:newton}
\end{equation}
We therefore solve Newton's equation in the Earth's rest frame to obtain each observer's worldline $\bm{x}(t)$, and feed the resulting $\Phi(\bm{x}(t))$ and $v^2(t)$ into Eq.~\eqref{eq:weakrate}. Post-Newtonian corrections to the trajectory are of order $GM/(Rc^2)\sim 7\times 10^{-10}$ and enter $\tau$ only at $O(c^{-4})$, well below the differences of interest.

\subsection{Earth model and Alice's trajectory}

Inside a uniform-density sphere of radius $R$ and mass $M$, the potential is
\begin{equation}
    \Phi_{\text{int}}(r) = -\frac{GM}{2R^3}\!\left(3R^2 - r^2\right),
    \label{eq:phiinside}
\end{equation}
and the gravitational force is linear: $\bm{g}(r) = -(GM/R^3)\,\bm{r}$. Equation~\eqref{eq:phiinside} is the exact Newtonian interior potential---readers can recover it as the work per unit mass needed to bring a test particle in from infinity---and is equally the leading $O(1/c^2)$ input to the metric~\eqref{eq:weakmetric}. Alice, released from rest at $r = R$, executes simple harmonic motion $r(t) = R\cos(\omega t)$ with $\omega = \sqrt{GM/R^3}$, where here $r$ is the signed coordinate along the tunnel (negative once she passes the centre). Her round-trip time is the gravity-train period~\cite{Wheeler1990}
\begin{equation}
    T = \frac{2\pi}{\omega} \approx 5060\;\text{s} \approx 84.3\;\text{min}\,,
    \label{eq:period}
\end{equation}
the same as the orbital period of a satellite grazing the surface, since both motions share the frequency~$\omega$\footnote{This invites a useful comparison. A satellite in a circular, ``grazing'' orbit at $r=R$ traces a timelike geodesic that connects the same two events ``passing through the release point'' separated by one coordinate period $T$. But its elapsed proper time is not the same as Bob's. In the weak-field approximation~\eqref{eq:weakrate}, the satellite has the same potential $\Phi(R)=-GM/R$ as Bob but nonzero speed $v^2=GM/R$, so
\begin{equation}
    \tau_{\text{sat}} \simeq T\!\left(1-\frac{GM}{Rc^2}-\frac{1}{2}\frac{GM}{Rc^2}\right)=T\!\left(1-\frac{3}{2}\frac{GM}{Rc^2}\right),
\end{equation}
and therefore $\tau_B-\tau_{\text{sat}}\simeq T\,GM/(2Rc^2)\approx 1.8\,\mu\text{s}$ for Earth. (In Schwarzschild spacetime, the same inequality appears as $d\tau/dt=\sqrt{1-3GM/(Rc^2)}$ for a circular geodesic versus $d\tau/dt=\sqrt{1-2GM/(Rc^2)}$ for a static clock at the same radius.) This is another instance of the paper's moral: coordinate-period coincidences do not fix the proper time; one must compute $\tau$ using Eq.~\eqref{eq:weakrate}.}.

\subsection{Bob and Alice compared}

Bob stays at $r = R$ with $v = 0$, so only the gravitational term in Eq.~\eqref{eq:weakrate} survives; integrating $d\tau/dt$ over the coordinate interval $T$ gives his elapsed proper time
\begin{equation}
    \tau_B \simeq T\!\left(1 - \frac{GM}{Rc^2}\right).
    \label{eq:tauB}
\end{equation}
For Alice's SHM trajectory, the time-averages over a full period, $\langle f\rangle \equiv T^{-1}\int_0^T f\,dt$, are
\begin{equation}
    \langle r^2 \rangle = \frac{R^2}{2}\,,\qquad \langle v^2 \rangle = \frac{\omega^2 R^2}{2} = \frac{GM}{2R}\,,
    \label{eq:averages}
\end{equation}
and the averaged potential along her worldline is
\begin{equation}
    \langle \Phi \rangle = -\frac{GM}{2R^3}\!\left(3R^2 - \langle r^2 \rangle\right) = -\frac{5}{4}\frac{GM}{R}\,.
    \label{eq:avgPhi}
\end{equation}
Since $d\tau/dt$ in Eq.~\eqref{eq:weakrate} is linear in $\Phi$ and quadratic in $v$, integrating it along Alice's worldline gives, exactly at this order in $1/c^2$,
\begin{equation}
    \tau_A = \int_0^T \frac{d\tau}{dt}\,dt \simeq T\!\left(1 + \frac{\langle \Phi \rangle}{c^2} - \frac{\langle v^2 \rangle}{2c^2}\right) = T\!\left(1 - \frac{3}{2}\frac{GM}{Rc^2}\right).
    \label{eq:tauA}
\end{equation}
The averages are not an approximation: for the harmonic-oscillator trajectory and the harmonic potential, the time-integrals of $v^2$ and $\Phi$ are evaluated in closed form, and the only approximation in Eq.~\eqref{eq:tauA} is the weak-field expansion of $d\tau/dt$ itself.
Therefore
\begin{equation}
    \tau_B - \tau_A \simeq \frac{T}{2}\,\frac{GM}{Rc^2} \approx 1.8\;\mu\text{s}\,.
    \label{eq:deltatau}
\end{equation}
Alice is geodesic, yet she ages less than Bob. The deficit splits equally between the gravitational and kinematic terms---a consequence of the virial theorem for the harmonic potential.

\subsection{Charlie: a geodesic that does maximise}\label{sec:charlie}

Charlie departs from the same surface event but follows a radial free-fall trajectory in the exterior vacuum, reaching apogee $r_{\max}$ and returning at $t = T$.

In the Newtonian exterior potential $\Phi(r) = -GM/r$, energy conservation gives
\begin{equation}
    v(r) = \sqrt{2GM\!\left(\frac{1}{r} - \frac{1}{r_{\max}}\right)}\,,
    \label{eq:vradial}
\end{equation}
and the Newtonian flight time is
\begin{equation}
    T = 2\sqrt{\frac{r_{\max}^3}{2GM}}\left[\arccos\sqrt{\frac{R}{r_{\max}}} + \sqrt{\frac{R}{r_{\max}}\!\left(1 - \frac{R}{r_{\max}}\right)}\right],
    \label{eq:Tcharlie}
\end{equation}
which implicitly fixes $r_{\max}$ once $T$ is given.\footnote{Both this integral and the integral $J$ in Eq.~\eqref{eq:Jclosed} below are reduced to elementary form by the substitution $r = r_{\max}\sin^2\theta$, which sends the turning points $r=R$ and $r=r_{\max}$ to fixed limits in $\theta$. The closed forms then follow from $\frac{d}{du}\!\left(\arccos\sqrt{u} + \sqrt{u(1-u)}\right) = -\sqrt{u/(1-u)}$, evaluated with $u = R/r_{\max}$.} Using Eq.~\eqref{eq:vradial} to eliminate $v^2$ in the $O(c^{-2})$ correction (with $\Phi(r) = -GM/r$ the exterior potential, since Charlie stays outside the Earth throughout),
\begin{equation}
    \Phi(r) - \frac{v(r)^2}{2} = GM\!\left(\frac{1}{r_{\max}} - \frac{2}{r}\right),
\end{equation}
so that
\begin{equation}
    \tau_C \simeq T + \frac{GM}{c^2}\,J(R, r_{\max})\,,
    \label{eq:tauCdef}
\end{equation}
with $J$ the integral
\begin{equation}
    J(R, r_{\max}) = 2\int_R^{r_{\max}} \left(\frac{1}{r_{\max}} - \frac{2}{r}\right)\frac{dr}{v(r)}\,.
\end{equation}
Evaluating in closed form:
\begin{equation}
    J(R, r_{\max}) = \frac{\sqrt{2}}{\sqrt{GM\,r_{\max}}}\left[\sqrt{R(r_{\max} - R)} - 3r_{\max}\arctan\sqrt{\frac{r_{\max} - R}{R}}\right].
    \label{eq:Jclosed}
\end{equation}
Combining Eqs.~\eqref{eq:tauB} and~\eqref{eq:tauCdef}, the proper-time difference is
\begin{equation}
    \tau_C - \tau_B = \frac{GM}{c^2}\!\left[J(R, r_{\max}) + \frac{T}{R}\right].
    \label{eq:tauCminusB}
\end{equation}
For $T \approx 5060\;\text{s}$, Eq.~\eqref{eq:Tcharlie} gives $r_{\max} \approx 1.42 \times 10^7\;\text{m}$ (altitude $\approx 1.23\,R$), and
\begin{equation}
    \tau_C - \tau_B \approx 1.1\;\mu\text{s} > 0\,.
    \label{eq:tauCminusBnum}
\end{equation}
Charlie, unlike Alice, does conform to the geodesic slogan: he ages more than the accelerated Bob. The difference between the two geodesic observers is that Charlie's worldline, as we shall now see, does not pass through a conjugate point.

\section{Why Alice's geodesic is not a maximiser}\label{sec:conjugate}

\subsection{Conjugate points}

Let $\gamma$ be a timelike geodesic from $p$ to $q$, and let $\gamma_s$ be a one-parameter family of timelike geodesics with $\gamma_0 = \gamma$ and all members starting at $p$. The deviation vector
\begin{equation*}
    \eta^\mu(\tau) = \left.\frac{\partial \gamma_s^\mu(\tau)}{\partial s}\right|_{s=0},
\end{equation*}
evaluated at the same proper-time parameter $\tau$ on each curve, gives the infinitesimal separation between $\gamma$ and a neighbouring member of the family at corresponding events. This $\eta^\mu$ is a \emph{Jacobi field} along $\gamma$, and satisfies the geodesic deviation equation
\begin{equation}
    \frac{D^2 \eta^\mu}{d\tau^2} + R^\mu{}_{\nu\alpha\beta}\,u^\nu \eta^\alpha u^\beta = 0,
    \label{eq:jacobi}
\end{equation}
where $D/d\tau$ is the covariant derivative along $\gamma$, $R^\mu{}_{\nu\alpha\beta}$ is the Riemann curvature tensor, and $u^\mu$ is the unit tangent to $\gamma$. If a nontrivial Jacobi field vanishes at both $p$ and at some later point $q$ on $\gamma$ --- so that the family of geodesics from $p$ refocuses at $q$ to first order in $s$ --- then $q$ is \emph{conjugate} to $p$ along $\gamma$. The standard result (Wald~\cite{Wald1984}, Theorem 9.5.1; see also Refs.~\cite{Hawking1973,Beem1996}) is that a timelike geodesic from $p$ to $q$ locally maximises proper time among nearby timelike curves connecting $p$ and $q$ if and only if no point conjugate to $p$ lies strictly between $p$ and $q$ along the geodesic.

\subsection{The antipode as conjugate point}

For the weak-field metric~\eqref{eq:weakmetric} in the slow-motion limit, the only Riemann tensor components contributing to Eq.~\eqref{eq:jacobi} are the tidal tensor $R^i{}_{0j0} = \partial_i \partial_j \Phi$. Inside the uniform-density sphere, the potential~\eqref{eq:phiinside} gives $\partial_i \partial_j \Phi_{\text{int}} = (GM/R^3)\,\delta_{ij} = \omega^2\,\delta_{ij}$, and the geodesic-deviation equation becomes
\begin{equation}
    \ddot{\eta}^i + \omega^2 \eta^i = 0,
    \label{eq:jacobiSHM}
\end{equation}
which is just the linearised version of the Newtonian SHM equation $\ddot{x}^i = -\omega^2 x^i$ that nearby test particles inside the Earth obey. So the Jacobi field along Alice's worldline is, to leading order, the deviation between her trajectory and any nearby Newtonian SHM trajectory.

Choose a transverse perturbation. Pair Alice's worldline $\bm{x}(t) = R\cos(\omega t)\,\hat{\bm{x}}$ with a nearby geodesic departing from the same surface event with a small \emph{transverse} initial-velocity component $\epsilon\,\hat{\bm{y}}$. Both worldlines remain inside the uniform-density sphere for sufficiently small $\epsilon$, so the linear analysis stays within the model.\footnote{A radial perturbation of the initial velocity also produces a Jacobi field with the same zeros, but the perturbed trajectory acquires amplitude $\sqrt{R^2+(\epsilon/\omega)^2} > R$ and so leaves the uniform-density interior. The transverse perturbation avoids this complication.} The perturbed trajectory is
\begin{equation}
    \bm{x}_\epsilon(t) = R\cos(\omega t)\,\hat{\bm{x}} + \frac{\epsilon}{\omega}\sin(\omega t)\,\hat{\bm{y}},
\end{equation}
and the Jacobi field
\begin{equation}
    \bm{\eta}(t) = \left.\frac{\partial \bm{x}_\epsilon}{\partial \epsilon}\right|_{\epsilon=0} = \frac{1}{\omega}\sin(\omega t)\,\hat{\bm{y}}
\end{equation}
indeed solves Eq.~\eqref{eq:jacobiSHM} and vanishes at $t = 0$ and at $t = T/2 = \pi/\omega$---when Alice reaches the antipodal surface. By spherical symmetry, all transverse perturbations produce conjugate points at the same location. (The Jacobi equation~\eqref{eq:jacobi} is written in proper time $\tau$; using coordinate time $t$ here, with $t = \tau + O(c^{-2})$ along Alice, shifts the focusing time and location only at higher order, so the antipode is a conjugate point to leading order in the weak-field expansion.)

The physical content is simply the isochrony of the harmonic oscillator: nearby Newtonian trajectories starting at the same point with slightly different initial velocities all reconverge at the antipodal surface after exactly half a period. That reconvergence is the conjugate point. Past it, Alice's worldline remains stationary for the proper-time functional, as every geodesic does; what it loses is local maximality. In the language of Morse theory~\cite{Milnor1963}, past the antipode the index form is no longer positive---equivalently, the second variation of the proper-time functional acquires \emph{positive} directions---so the stationary point is a saddle rather than a maximum.\footnote{With the index form $I[\bm\eta]=\int_0^T\!\big(|\dot{\bm\eta}|^2-\omega^2|\bm\eta|^2\big)\,dt$, one has $\delta^2\tau = -I/c^2$, so local proper-time maximality corresponds to $I$ positive definite. By the Morse index theorem the number of negative directions of $I$ equals the conjugate-point multiplicity; by spherical symmetry both directions transverse to Alice's radial motion refocus at the antipode, so the multiplicity is two (it reduces to one if attention is restricted to a single transverse plane).} There then exist \emph{nearby} timelike curves---small deformations of Alice's worldline with the same endpoints---of strictly greater elapsed proper time. Bob's worldline is not one of these (it never enters the Earth); it is a separate, explicit comparison curve, computed above, that also out-ages Alice.

Charlie's geodesic, by contrast, stays in the exterior vacuum, where $\Phi_{\text{ext}}=-GM/r$. The radial tide $\partial_r^2 \Phi_{\text{ext}} = -2GM/r^3$ is defocusing; the transverse tide is focusing, and a transverse Jacobi field along his worldline obeys
\begin{equation*}
    \ddot{\eta}^i = -\frac{GM}{r(t)^3}\,\eta^i,
\end{equation*}
with effective frequency $\omega_\perp(t) = \sqrt{GM/r(t)^3} \leq \omega$, since $r(t) \geq R$; Charlie's transverse field thus oscillates no faster than Alice's. In fact this focusing equation can be solved in closed form along the radial trajectory. One solution is $\eta(t) = r(t)$ itself, since $\ddot{r} = -GM/r^2$ implies $\ddot{r} = -(GM/r^3)\,r$; reduction of order then gives the transverse Jacobi field that vanishes at departure,
\begin{equation*}
    \eta_\perp(t) = r(t)\int_0^t \frac{dt'}{r(t')^2}\,,
\end{equation*}
which is strictly positive for all $0 < t \leq T$ because $r(t) > 0$ throughout. The radial mode obeys the defocusing equation $\ddot{\eta} = +2GM/r^3\,\eta$ and, starting from $\eta(0) = 0$ with $\dot\eta(0) > 0$, likewise never returns to zero. Hence no conjugate point lies between Charlie's departure and reunion, and the local-maximality theorem applies to him.

\subsection{A Riemannian picture}\label{sec:sphere}

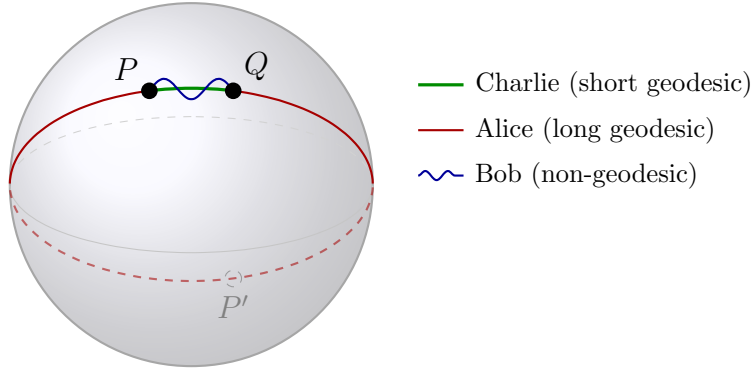
\begin{figure}[t]
\centering
\begin{tikzpicture}[scale=2.4]
  \shade[ball color=blue!8!white,opacity=0.35] (0,0) circle (1);
  \draw[thick,gray!70] (0,0) circle (1);
  \draw[gray!45] plot coordinates {(-1.000,-0.004) (-0.998,-0.021) (-0.995,-0.038) (-0.989,-0.055) (-0.982,-0.072) (-0.972,-0.088) (-0.960,-0.104) (-0.947,-0.121) (-0.931,-0.136) (-0.914,-0.152) (-0.895,-0.167) (-0.873,-0.182) (-0.851,-0.197) (-0.826,-0.211) (-0.800,-0.225) (-0.772,-0.238) (-0.742,-0.251) (-0.711,-0.263) (-0.679,-0.275) (-0.645,-0.286) (-0.610,-0.297) (-0.573,-0.307) (-0.535,-0.316) (-0.497,-0.325) (-0.457,-0.333) (-0.416,-0.341) (-0.375,-0.347) (-0.333,-0.353) (-0.290,-0.359) (-0.246,-0.363) (-0.202,-0.367) (-0.158,-0.370) (-0.113,-0.372) (-0.068,-0.374) (-0.023,-0.375) (0.023,-0.375) (0.068,-0.374) (0.113,-0.372) (0.158,-0.370) (0.202,-0.367) (0.246,-0.363) (0.290,-0.359) (0.333,-0.353) (0.375,-0.347) (0.416,-0.341) (0.457,-0.333) (0.497,-0.325) (0.535,-0.316) (0.573,-0.307) (0.610,-0.297) (0.645,-0.286) (0.679,-0.275) (0.711,-0.263) (0.742,-0.251) (0.772,-0.238) (0.800,-0.225) (0.826,-0.211) (0.851,-0.197) (0.873,-0.182) (0.895,-0.167) (0.914,-0.152) (0.931,-0.136) (0.947,-0.121) (0.960,-0.104) (0.972,-0.088) (0.982,-0.072) (0.989,-0.055) (0.995,-0.038) (0.998,-0.021) (1.000,-0.004)};
  \draw[gray!35,dashed] plot coordinates {(-0.000,0.375) (-0.045,0.374) (-0.090,0.373) (-0.135,0.371) (-0.180,0.368) (-0.224,0.365) (-0.268,0.361) (-0.311,0.356) (-0.354,0.350) (-0.396,0.344) (-0.437,0.337) (-0.477,0.329) (-0.516,0.321) (-0.554,0.312) (-0.591,0.302) (-0.627,0.292) (-0.662,0.281) (-0.695,0.269) (-0.727,0.257) (-0.757,0.245) (-0.786,0.232) (-0.813,0.218) (-0.838,0.204) (-0.862,0.190) (-0.884,0.175) (-0.904,0.160) (-0.923,0.144) (-0.939,0.129) (-0.954,0.113) (-0.966,0.096) (-0.977,0.080) (-0.986,0.063) (-0.992,0.046) (-0.997,0.030) (-0.999,0.013)};
  \draw[gray!35,dashed] plot coordinates {(0.999,0.013) (0.997,0.030) (0.992,0.046) (0.986,0.063) (0.977,0.080) (0.966,0.096) (0.954,0.113) (0.939,0.129) (0.923,0.144) (0.904,0.160) (0.884,0.175) (0.862,0.190) (0.838,0.204) (0.813,0.218) (0.786,0.232) (0.757,0.245) (0.727,0.257) (0.695,0.269) (0.662,0.281) (0.627,0.292) (0.591,0.302) (0.554,0.312) (0.516,0.321) (0.477,0.329) (0.437,0.337) (0.396,0.344) (0.354,0.350) (0.311,0.356) (0.268,0.361) (0.224,0.365) (0.180,0.368) (0.135,0.371) (0.090,0.373) (0.045,0.374) (0.000,0.375)};
  \draw[red!65!black, thick, dashed, opacity=0.55] plot coordinates {(1.000,-0.015) (0.998,-0.036) (0.994,-0.057) (0.989,-0.077) (0.983,-0.098) (0.975,-0.118) (0.966,-0.138) (0.955,-0.158) (0.942,-0.178) (0.929,-0.197) (0.914,-0.216) (0.897,-0.235) (0.879,-0.253) (0.860,-0.271) (0.839,-0.289) (0.817,-0.306) (0.794,-0.323) (0.770,-0.339) (0.744,-0.355) (0.718,-0.370) (0.690,-0.384) (0.661,-0.398) (0.631,-0.412) (0.601,-0.425) (0.569,-0.437) (0.536,-0.448) (0.503,-0.459) (0.469,-0.469) (0.434,-0.478) (0.399,-0.487) (0.362,-0.495) (0.326,-0.502) (0.289,-0.509) (0.251,-0.514) (0.213,-0.519) (0.175,-0.523) (0.136,-0.526) (0.097,-0.529) (0.059,-0.530) (0.020,-0.531) (-0.020,-0.531) (-0.059,-0.530) (-0.097,-0.529) (-0.136,-0.526) (-0.175,-0.523) (-0.213,-0.519) (-0.251,-0.514) (-0.289,-0.509) (-0.326,-0.502) (-0.362,-0.495) (-0.399,-0.487) (-0.434,-0.478) (-0.469,-0.469) (-0.503,-0.459) (-0.536,-0.448) (-0.569,-0.437) (-0.601,-0.425) (-0.631,-0.412) (-0.661,-0.398) (-0.690,-0.384) (-0.718,-0.370) (-0.744,-0.355) (-0.770,-0.339) (-0.794,-0.323) (-0.817,-0.306) (-0.839,-0.289) (-0.860,-0.271) (-0.879,-0.253) (-0.897,-0.235) (-0.914,-0.216) (-0.929,-0.197) (-0.942,-0.178) (-0.955,-0.158) (-0.966,-0.138) (-0.975,-0.118) (-0.983,-0.098) (-0.989,-0.077) (-0.994,-0.057) (-0.998,-0.036) (-1.000,-0.015)};
  \draw[red!65!black, thick] plot coordinates {(0.231,0.517) (0.268,0.512) (0.306,0.506) (0.343,0.499) (0.379,0.491) (0.415,0.483) (0.450,0.474) (0.485,0.465) (0.518,0.454) (0.551,0.443) (0.584,0.431) (0.615,0.419) (0.645,0.406) (0.675,0.392) (0.703,0.378) (0.730,0.363) (0.756,0.348) (0.781,0.332) (0.805,0.315) (0.827,0.298) (0.849,0.281) (0.869,0.263) (0.887,0.245) (0.905,0.226) (0.921,0.207) (0.935,0.188) (0.948,0.169) (0.960,0.149) (0.970,0.129) (0.979,0.109) (0.986,0.088) (0.992,0.068) (0.996,0.047) (0.999,0.026) (1.000,0.006)};
  \draw[red!65!black, thick] plot coordinates {(-1.000,0.006) (-0.999,0.026) (-0.996,0.047) (-0.992,0.068) (-0.986,0.088) (-0.979,0.109) (-0.970,0.129) (-0.960,0.149) (-0.948,0.169) (-0.935,0.188) (-0.921,0.207) (-0.905,0.226) (-0.887,0.245) (-0.869,0.263) (-0.849,0.281) (-0.827,0.298) (-0.805,0.315) (-0.781,0.332) (-0.756,0.348) (-0.730,0.363) (-0.703,0.378) (-0.675,0.392) (-0.645,0.406) (-0.615,0.419) (-0.584,0.431) (-0.551,0.443) (-0.518,0.454) (-0.485,0.465) (-0.450,0.474) (-0.415,0.483) (-0.379,0.491) (-0.343,0.499) (-0.306,0.506) (-0.268,0.512) (-0.231,0.517)};
  \draw[green!55!black, very thick] plot coordinates {(-0.231,0.517) (-0.209,0.519) (-0.187,0.522) (-0.165,0.524) (-0.144,0.526) (-0.122,0.527) (-0.100,0.528) (-0.077,0.530) (-0.055,0.530) (-0.033,0.531) (-0.011,0.531) (0.011,0.531) (0.033,0.531) (0.055,0.530) (0.077,0.530) (0.100,0.528) (0.122,0.527) (0.144,0.526) (0.165,0.524) (0.187,0.522) (0.209,0.519) (0.231,0.517)};
  \draw[blue!60!black, thick] plot coordinates {(-0.231,0.517) (-0.223,0.527) (-0.215,0.537) (-0.207,0.547) (-0.200,0.555) (-0.192,0.563) (-0.184,0.570) (-0.176,0.575) (-0.168,0.579) (-0.161,0.582) (-0.153,0.583) (-0.145,0.582) (-0.137,0.581) (-0.130,0.577) (-0.122,0.573) (-0.114,0.567) (-0.106,0.560) (-0.098,0.553) (-0.091,0.544) (-0.083,0.536) (-0.075,0.526) (-0.067,0.517) (-0.059,0.508) (-0.051,0.500) (-0.043,0.492) (-0.035,0.485) (-0.028,0.480) (-0.020,0.475) (-0.012,0.472) (-0.004,0.471) (0.004,0.471) (0.012,0.472) (0.020,0.475) (0.028,0.480) (0.035,0.485) (0.043,0.492) (0.051,0.500) (0.059,0.508) (0.067,0.517) (0.075,0.526) (0.083,0.536) (0.091,0.544) (0.098,0.553) (0.106,0.560) (0.114,0.567) (0.122,0.573) (0.130,0.577) (0.137,0.581) (0.145,0.582) (0.153,0.583) (0.161,0.582) (0.168,0.579) (0.176,0.575) (0.184,0.570) (0.192,0.563) (0.200,0.555) (0.207,0.547) (0.215,0.537) (0.223,0.527) (0.231,0.517)};
  \fill (-0.231,0.517) circle (1.3pt) node[above left] {$P$};
  \fill (0.231,0.517) circle (1.3pt) node[above right] {$Q$};
  \draw[gray,dashed,opacity=0.7] (0.231,-0.517) circle (1.3pt);
  \node[gray,below] at (0.231,-0.547) {$P'$};

  \draw[green!55!black, very thick] (1.25,0.55) -- (1.5,0.55) node[right, black]{\footnotesize Charlie (short geodesic)};
  \draw[red!65!black, thick] (1.25,0.30) -- (1.5,0.30) node[right, black]{\footnotesize Alice (long geodesic)};
  \draw[blue!60!black, thick, decorate, decoration={snake, amplitude=0.6mm, segment length=3mm}] (1.25,0.05) -- (1.5,0.05) node[right, black]{\footnotesize Bob (non-geodesic)};
\end{tikzpicture}
\caption{Riemannian analogy on $S^2$. The short great-circle arc (green, ``Charlie'') minimises length and does not pass through a conjugate point. The long arc (red, ``Alice'') passes through the antipode $P'$---a conjugate point---and is no longer length-minimising. The squiggly non-geodesic curve (blue, ``Bob'') goes directly from $P$ to $Q$ and is shorter than Alice's long geodesic despite not being a geodesic itself. In the Lorentzian boomerang, replace ``shorter'' by ``longer in proper time'' and ``minimising'' by ``maximising.''}
\label{fig:sphere}
\end{figure}

A useful analogy lives on the 2-sphere $S^2$ (Fig.~\ref{fig:sphere}). On the sphere, great circles are geodesics and they minimise arc length---but only up to the first conjugate point, which here is the antipodal point. (On the sphere this conjugate point coincides with the \emph{cut locus}; in more general Riemannian geometry the two notions need not coincide.)

Take two nearby points $P$ and $Q$. The short great-circle arc connecting them minimises length. The long arc, which goes the wrong way around through the antipode $P'$, does not: having passed through a conjugate point, it is no longer length-minimising---though, being a geodesic, it remains a stationary point of the length functional. Indeed, a non-geodesic curve that simply cuts across from $P$ to $Q$ without following any great circle will typically be shorter than the long arc, even though it is not a geodesic.

The dictionary with our three observers writes itself: Charlie is the short arc, Alice the long arc, Bob the squiggly shortcut. Once Alice's geodesic has passed through the conjugate point, there is nothing privileged about its length, and Bob's non-geodesic worldline---which, like Bob himself, stays near the surface rather than plunging through---can and does beat it.

There are two disanalogies worth flagging. First, the sense of the extremum reverses: minimisation of length on the sphere becomes maximisation of proper time in spacetime. Second---and this is the more interesting point---on the sphere it is tempting to read the antipodal conjugate point as a feature of the closed, compact geometry; but the refocusing is in fact a matter of \emph{curvature}, the sphere's positive Gaussian curvature focusing nearby geodesics. The boomerang makes this unmistakable: the same curvature-focusing produces a conjugate point on a topologically trivial $\mathbb{R}^4$, sourced by an ordinary amount of matter, with no spatial compactification at all. This same focusing mechanism, described by the Raychaudhuri equation~\cite{Hawking1973}, underpins the Penrose--Hawking singularity theorems~\cite{Penrose1965,Hawking1966}.

\section{Relation to other twin paradoxes}\label{sec:discussion}

In the standard twin paradox of special relativity, the geodesic (inertial) worldline connecting two events in Minkowski space is unique, so there is no room for confusion: the inertial twin always ages the most. Curvature changes this. In a curved spacetime, multiple geodesics can connect the same pair of events, and they need not all be proper-time maximisers. Our thought experiment is a concrete instance of this.

A related phenomenon occurs in flat spacetime with compact spatial topology~\cite{Barrow2001,Uzan2002}. If one dimension is rolled up into a cylinder, multiple inertial worldlines with different winding numbers connect any two events, and neither twin need accelerate to produce an age difference. That case eliminates acceleration but requires non-trivial topology. The boomeranging-twins example eliminates both: neither Alice nor Charlie accelerates, and the spacetime is topologically $\mathbb{R}^4$. What it requires instead is curvature sufficient to produce a conjugate point---which is to say, a rather ordinary amount of matter.

It is perhaps worth noting how the three pairwise comparisons relate to the standard intuition. Bob versus Charlie is the textbook case: the geodesic (Charlie) ages more than the accelerated observer (Bob). Alice versus Bob is the surprise: the geodesic (Alice) ages \emph{less} than the accelerated observer. Alice versus Charlie completes the picture: two geodesics, same endpoints, different proper times---only the one free of intermediate conjugate points remains even a local proper-time maximiser, and here it is also the longer of the two. The same accelerating observer, Bob, sits on opposite sides of the comparison depending on which geodesic he is paired with. That is what makes the GR slogan unreliable as a universal heuristic, and what makes the pre-geometric bookkeeping of Eq.~\eqref{eq:weakrate} the more trustworthy guide \emph{here}.

\section{Conclusion}\label{sec:conclusion}

Bell showed that, in certain cases, pre-relativistic intuitions about Lorentz contraction are more reliable than the principles of special relativity for getting the right answer to a genuinely relativistic problem~\cite{Bell1976}. The boomeranging twins make the analogous point one level up: there are simple problems, involving genuinely general-relativistic effects, where the older pre-geometric heuristic---velocity and gravitational time dilation, combined in a single frame---gets the answer right at once, while the GR slogan ``geodesics maximise proper time'' misleads. Neither heuristic dominates in general: for problems where the geometry itself is the object of interest, or where strong-field effects matter, the geometric machinery is essential and the pre-geometric bookkeeping would be cumbersome or simply unavailable. The point of the example is rather that the slogan, treated as a universal rule of thumb, can mislead even where the pre-geometric calculation is short.

The slogan fails because it elides a geometric qualification: the maximality of a timelike geodesic is local, and lapses once the geodesic passes a conjugate point. In the thought experiment, that conjugate point is as concrete as it could be---it is the far side of the Earth, where the harmonic oscillator's isochrony refocuses neighbouring geodesics. The example requires no exotic ingredients: just a uniform-density sphere, a vacuum tube, and the weak-field limit. As a classroom tool it can serve double duty, both as a cautionary tale about slogans and as a gateway to conjugate points, the Raychaudhuri equation, and the singularity theorems.

\section*{Acknowledgments}
I thank Christopher Timpson and Oliver Pooley for discussions that prompted the formulation of this puzzle. The Riemannian analogy on the 2-sphere is due to Pooley. The authors gratefully acknowledge funding from the European Research Council, Grant 101088528 COGY.

\end{document}